\documentclass[english,conference, twocolumn]{IEEEtran}
\usepackage[dvips]{color}
\usepackage[T1]{fontenc}
\usepackage{amsmath}
\usepackage{amssymb,latexsym}
\usepackage{epsfig}
\usepackage{graphicx}

\makeatletter
\newtheorem{definitn}{Definition}

\usepackage[dvips]{color}

\makeatother

\usepackage{babel}

\begin{document}

\title{Transmission Capacities for Overlaid Wireless Ad Hoc Networks with
Outage Constraints}

\author{\IEEEauthorblockN{Changchuan Yin, Long Gao, Tie Liu, and Shuguang Cui}
\IEEEauthorblockA{Department of Electrical and Computer Engineering\\
Texas A\&M University, College Station, TX 77843\\
 Email: ccyin@ieee.org, \{lgao, tieliu, cui\}@ece.tamu.edu}}

\maketitle
\begin{abstract}
We study the transmission capacities of two coexisting wireless
networks (a primary network vs. a secondary network) that operate in
the same geographic region and share the same spectrum. We define
transmission capacity as the product among the density of
transmissions, the transmission rate, and the successful
transmission probability (1 minus the outage probability). The
primary (PR) network has a higher priority to access the spectrum
without particular considerations for the secondary (SR) network,
where the SR network limits its interference to the PR network by
carefully controlling the density of its transmitters. Assuming that
the nodes are distributed according to Poisson point processes and
the two networks use different transmission ranges, we quantify the
transmission capacities for both of these two networks and discuss
their tradeoff based on asymptotic analyses. Our results show that
if the PR network permits a small increase of its outage
probability, the sum transmission capacity of the two networks
(i.e., the overall spectrum efficiency per unit area) will be
boosted significantly over that of a single network.
\end{abstract}

\section{Introduction}

Initiated by the seminal work of Gupta and
Kumar~\cite{Capacity:Gupta}, the studies for understanding the
capacities of wireless ad hoc networks have made great progresses.
Considering $n$ nodes that are randomly distributed in a unit area
and grouped independently into one-to-one source-destination (S-D)
pairs, Gupta and Kumar~\cite{Capacity:Gupta} showed that a typical
time-slotted multi-hop architecture with a common transmission range
and adjacent-neighbor communication can achieve a sum throughput
that scales as $\Theta\left(\sqrt{n/\log n}\right)$. By using
percolation theory, Franceschetti \emph{et~al.}~\cite{Gap:Fran}
showed that the $\Theta\left(\sqrt{n}\right)$ sum throughput scaling
is achievable. In~\cite{Mobility:Grossglauser}, Grossglauser and Tse
showed that by allowing the nodes to move independently and
uniformly, a constant throughput scaling $\Theta(1)$ per S-D pair
can be achieved. In~\cite{Baccelli:Aloha}, Baccelli \emph{et~al.}
proposed a multi-hop spatial reuse ALOHA protocol. By optimizing the
product between the number of simultaneous successful transmissions
per unit area and the average transmission range, they showed that
the transport capacity is proportional to the square root of the
node density, which achieves the upper bound of Gupta and
Kumar~\cite{Capacity:Gupta}. Weber \emph{et al}
in~\cite{Weber:Capacity} derived the upper and lower bounds on
transmission capacity of spread-spectrum wireless ad hoc networks,
where the transmission capacity is defined as the product between
the maximum density of successful transmissions and the
corresponding data rate, under a constraint on the outage
probability.

All the above results focus on the capacity of a single ad hoc
wireless network. In recent years, due to the scarcity and poor
utilization of spectrum, the regulation bodies are beginning to
consider the possibility of permitting secondary (SR) networks to
coexist with licensed primary (PR) networks, which is the main
driving force behind the cognitive radio
technology~\cite{Cognitive:Haykin}. In cognitive radio networks, the
PR users have a higher priority to access the spectrum and the SR
users need to operate conservatively such that their interference to
the PR users is limited below an ``acceptable level''. In this
overlaid regime, the capacity or throughput scaling laws for both of
the PR and SR networks are interesting
 problems. Recently, some preliminary works along this line appeared.
In~\cite{Scaling:Vu}, Vu~\emph{et~al.} considered the throughput
scaling law for a single-hop cognitive radio network, where a linear
scaling law is obtained for the SR network with an outage constraint
for the PR network. In~\cite{Jeon:Cognitive}, Jeon~\emph{et~al}.
considered a multi-hop cognitive network on top of a PR network and
assumed that the SR nodes know the location of each PR node. With an
elegant transmission scheme, they showed that by defining a
preservation region around each PR node, both networks can achieve
the same throughput scaling law as a stand-alone wireless network,
while the SR network may suffer from a finite outage probability.
In~\cite{Yin:Scaling}, Yin~\emph{et~al}. assumed that the SR nodes
only konw the locations of PR transmitters (TXs) and proposed a
transmission scheme to show that both networks can achieve the same
throughput scaling law as a stand-alone wireless network, with zero
outage.

In this paper, we study the coexistence of two ad hoc networks with
different transmission scales (power and/or transmission range)
based on the transmission capacity defined in~\cite{Weber:Capacity}.
We extend the definition of transmission capacity from a single
network to two overlaid networks. Different from the approaches
in~\cite{Scaling:Vu}, \cite{Jeon:Cognitive}, and~\cite{Yin:Scaling},
we resort to stochastic geometry tools to quantify the transmission
capacities for both the PR and SR networks without defining any
preservation regions. By considering the mutual interferences from
the two networks, we discuss the tradeoff of the transmission
capacities between them. The results show that if we permit a slight
increase over the outage probability of the PR network, the sum
transmission capacity (i.e., the overall spectrum efficiency per
unit area) of the overlaid networks will be boosted significantly
over that of a single network.

The rest of the paper is organized as follows. The network model and
symbol notations are described in Section II. The transmission
capacity for a single network case is analyzed in Section III. The
transmission capacities for the PR and SR networks and their
tradeoff are discussed in Section IV. The numerical results and
observations are given in Section V. Finally, Section VI summarizes
our conclusions.

\section{Network model and System Setup}

Consider the scenario where a network of PR nodes and a network of
SR nodes coexist in the same geographic region, and assume that the
PR network is the legacy network, which has a higher priority to
access the spectrum. The prerequisite condition for introducing a
new SR network into the territory of the PR network is that the
outage probability increment of the PR network is upper-limited by a
target constraint $\triangle\epsilon$, where $\triangle\epsilon$
usually takes a very small value.

We assume that at a certain time instance the distribution of PR TXs
follows a homogeneous Poisson point process (PPP) $\Pi_{0}$ of
density $\lambda_{0}$, and the distribution of SR TXs follows
another independent homogeneous PPP $\Pi_{1}$ of density
$\lambda_{1}$. Our goal is to evaluate the outage probability of the
PR network, $\mathcal{P}^{0}$, and that of the SR network,
$\mathcal{P}^{1}$, which are functions of the TX node densities
$\lambda_{0}$ and $\lambda_{1}$. The specific definitions of outage
probability will be given in Section III and Section IV. Similar to
that in~\cite{Weber:Capacity}, in order to evaluate the outage
probabilities, we condition on a typical PR (or SR) RX at the
origin, which yields the Palm distribution for PR (or SR) TXs.
Following the Slivnyak's theory in stochastic
geometry~\cite{Stoyan:Geometry}, these conditional distributions
also follow homogeneous PPPs with the corresponding densities (i.e.,
$\lambda_0$ and $\lambda_1$, respectively). Let $\left\{
X_{i}\in\mathbb{R}^{2},i\in\Pi_{0}\right\} $ and $\left\{
Y_{j}\in\mathbb{R}^{2},j\in\Pi_{1}\right\} $ denote the locations of
the PR TXs and the SR TXs, respectively, $\left|X_{i}\right|$ and
$\left|Y_{j}\right|$ denote
 the distances from PR
TX $i$ and SR TX $j$ to the origin, respectively. An attempted
transmission is successful if the received
signal-to-interference-plus-noise ratio (SINR) at the reference RX
is above a threshold, $\beta$; otherwise, the transmission fails,
i.e., an outage occurs. We use $\beta_{0}$ and $\beta_{1}$ to
represent the SINR thresholds for the PR network and the SR network,
respectively.

For simplicity, we limit our discussion to single-hop transmissions,
and assume that all PR TXs use the same transmission power
$\rho_{0}$, and all PR transmissions are over the same distance
$r_{0}$. Similarly, all SR TXs use the same transmission power
$\rho_{1}$ over the same transmission distance $r_{1}$. For the
wireless channel, we only consider the large-scale path-loss, and
ignore the effects of shadowing and small-scale multipath fading. As
such, the normalized channel power gain $g(d)$ is given
as\begin{equation}
g(d)=\frac{A}{d^{\alpha}},\label{Eq2}\end{equation} where $A$ is a
system-dependent constant, $d$ is the distance between the TX and
the corresponding RX, and $\alpha>2$ denotes the path-loss exponent.
In the following discussion, we normalize $A$ to be unity for
simplicity. The ambient noise is assumed to be additive white
Gaussian noise (AWGN) with an average power $\eta$. We assume that
all the PR TXs and the SR TXs use the same spectrum with bandwidth
normalized to be unity.

As in~\cite{Weber:Capacity}, we define transmission capacity as
follows.
\begin{definitn}
\emph{Transmission capacity} $C^{\epsilon}$ of a randomly-deployed
wireless network is defined as the product among the maximum density
$\lambda^{\epsilon}$ of transmissions, the common transmission data
rate $R$, and $(1-\epsilon)$ with $\epsilon$ an asymptotically small
outage probability. Therefore, we have\begin{equation}
C^{\epsilon}=R\lambda^{\epsilon}(1-\epsilon).\label{eq:Transmission_capacity}\end{equation}
As noted in\emph{~}\cite{Weber:Capacity}, $C^{\epsilon}$ also
represents the unit-area spectral efficiency of the successful
transmissions.
\end{definitn}

\section{Asymptotic Analysis of the Transmission Capacity: Single Network Case}
In this section, we derive the asymptotic result (asymptotic over
vanishingly-small outage probability values) for the transmission
capacity of the PR network when the SR network is absent. As an
example, we focus on the case when the path-loss exponent
$\alpha=4$, over which we build an asymptotic analysis framework
that is useful for the future study over the cases of general
$\alpha$ values.

When the SR network is absent, denote the target outage probability
of the PR network over per-link SINR as $\epsilon_0$. Then we have
\begin{equation}
\mathcal{P}^{0}=\textrm{Prob}\left(\frac{\rho_{0}r_{0}^{-\alpha}}{\eta+{\displaystyle
\sum_{i\in\Pi_{0}}\rho_{0}|X_{i}|^{-\alpha}}}\leq\beta_{0}\right){\displaystyle
=\epsilon_{0}}.\label{eq:P_outage}\end{equation} Rewrite
(\ref{eq:P_outage}) as
\begin{equation}
\textrm{Prob}\left({\displaystyle X}\geq T_{0}\right){\displaystyle
=\epsilon_{0}},\label{eq:Pout_s1}\end{equation} where
$X=\sum_{i\in\Pi_{0}}\rho_{0}|X_{i}|^{-\alpha}$ and
$T_{0}=\frac{\rho_{0}r_{0}^{-\alpha}}{\beta_{0}}-\eta.$ The moment
generating function (MGF) of $X$ is given
by~\cite{Venkataraman:Shot}
\begin{equation}
\Phi_{X}(s)=\exp\left[-\pi\lambda_{0}\rho_{0}^{\frac{2}{\alpha}}s^{\frac{2}{\alpha}}\Gamma\left(1-\frac{2}{\alpha}\right)\right].\end{equation}
When $\alpha=4$, we have
\begin{equation}
\Phi_{X}(s)=\exp\left[-\pi^{\frac{3}{2}}\lambda_{0}\rho_{0}^{\frac{1}{2}}s^{\frac{1}{2}}\right].\end{equation}
Via the inverse Laplace transform, we obtain the probability density
function (PDF) of $X$ as
\begin{equation}
f_{X}(x)=\frac{\pi}{2}\lambda_{0}\sqrt{\rho_{0}}x^{-\frac{3}{2}}\exp\left(-\frac{\pi^{3}}{4x}\lambda_{0}^{2}\rho_{0}\right),\end{equation}
and the corresponding cumulative density function (CDF) of $X$ as
\begin{eqnarray} F_{X}(x) & = &
2Q\left(\frac{\pi^{\frac{3}{2}}\lambda_{0}\sqrt{\rho_{0}}}{\sqrt{2x}}\right).\label{eq:CDF_P}\end{eqnarray}
From (\ref{eq:CDF_P}), we have
\begin{equation}
\textrm{Prob}\left({\displaystyle X}\geq
T_{0}\right)=1-2Q\left(\frac{\pi^{\frac{3}{2}}\lambda_{0}\sqrt{\rho_{0}}}{\sqrt{2T_{0}}}\right).\label{eq:PR_s_outage}\end{equation}
 Combined (\ref{eq:Pout_s1}) and (\ref{eq:PR_s_outage}), it is clear
 that the following condition has to be satisfied:
\begin{equation}\label{eq:Qfunc_PR}
Q\left(\frac{\pi^{\frac{3}{2}}\lambda_{0}\sqrt{\rho_{0}}}{\sqrt{2T_{0}}}\right)=\frac{1-\epsilon_{0}}{2}.\end{equation}
When $\epsilon_{0}\rightarrow0$ such that
$\frac{\pi^{\frac{3}{2}}\lambda_{0}\sqrt{\rho_{0}}}{\sqrt{2T_{0}}}\rightarrow0$,
 with Taylor series expansion, we obtain the maximum allowable value
 (via the monotonicity of the Q function)
of $\lambda_{0}$ asymptotically for $\alpha=4$ as
\begin{equation}
\lambda_{0}^{\epsilon_0}=\frac{\epsilon_{0}}{\pi}\left(\frac{T_{0}}{\rho_{0}}\right)^{\frac{1}{2}}=\frac{\epsilon_{0}}{\pi}\left(\frac{r_{0}^{-4}}{\beta_{0}}-\frac{\eta}{\rho_{0}}\right)^{\frac{1}{2}}.\label{eq:PRs_L0}\end{equation}
As we can see from (\ref{eq:PRs_L0}) that when the outage
probability $\epsilon_{0}$ is very small, the density of TXs is a
linear function of $\epsilon_{0}$. Therefore, the transmission
capacity of the PR network is given by
\begin{equation}
C_{0}^{\epsilon_{0}}=R_{0}\lambda_{0}^{\epsilon_0}(1-\epsilon_0),
\end{equation}
where $R_0$ is the data rate when the transmission between the TX
and its associated RX is successful, which is set to be same for all
the links.

\section{Asymptotic Analysis of the Transmission Capacity: Overlaid Network Case}

\subsection{Transmission Capacity of the PR Network}

When the SR network is present, it introduces interference to the PR
network and the outage probability of the PR network will be
increased. If we set the target outage probability increment of the
PR network as $\triangle\epsilon$, we have
\begin{eqnarray}
\mathcal{P}^{0}&=&\textrm{Prob}\left(\frac{\rho_{0}r_{0}^{-\alpha}}{\eta+{\displaystyle
\sum_{i\in\Pi_{0}}\rho_{0}|X_{i}|^{-\alpha}}+{\displaystyle
\sum_{j\in\Pi_{1}}\rho_{1}|Y_{j}|^{-\alpha}}}\leq\beta_{0}\right)\notag\\
&=&\epsilon_{0}+\triangle\epsilon.\label{eq:Bi_PR}
\end{eqnarray} With
$Y=\sum_{j\in\Pi_{1}}\rho_{1}|Y_{j}|^{-\alpha}$, (\ref{eq:Bi_PR})
can be rewritten as
\begin{equation} \textrm{Prob}\left(X+Y\geq
T_{0}\right){\displaystyle
=\epsilon_{0}}+\triangle\epsilon.\label{eq:Bi_PR_Err}\end{equation}
 The MGF of $Y$ is given by\begin{equation}
\Phi_{Y}(s)=\exp\left[-\pi\lambda_{1}\rho_{1}^{\frac{2}{\alpha}}s^{\frac{2}{\alpha}}\Gamma\left(1-\frac{2}{\alpha}\right)\right].\end{equation}
 Define $Z=X+Y$ such that the MGF of $Z$ is given by
\begin{eqnarray}
\nonumber \Phi_{Z}(s) &=& \Phi_{X}(s)\Phi_{Y}(s) \\
\nonumber             &=& \exp\left[-\pi
s^{\frac{2}{\alpha}}\Gamma\left(1-\frac{2}{\alpha}
\right)\left(\lambda_{0}\rho_{0}^{\frac{2}{\alpha}}+\lambda_{1}\rho_{1}^{\frac{2}{\alpha}}\right)\right].
\end{eqnarray}
For $\alpha=4$, we have
\begin{equation}
\Phi_{Z}(s)=\exp\left[-\pi^{\frac{3}{2}}
s^{\frac{1}{2}}\left(\lambda_{0}\sqrt{\rho_{0}}+\lambda_{1}\sqrt{\rho_{1}}\right)\right],
\end{equation}
 and the PDF of $Z$ is given by
{\setlength\arraycolsep{2pt}
\begin{eqnarray}
f_{Z}(z) &=& \frac{\pi}{2}\left(\lambda_{0}\sqrt{\rho_{0}}+\lambda_{1}\sqrt{\rho_{1}}\right)z^{-\frac{3}{2}}\nonumber\\
&&{}\times\exp\left[-\frac{\pi^{3}}{4z}\left(\lambda_{0}\sqrt{\rho_{0}}+\lambda_{1}\sqrt{\rho_{1}}\right)^{2}\right].
\label{eq:Bi_pdf_z}
\end{eqnarray}}
Applying (\ref{eq:Bi_pdf_z}) in (\ref{eq:Bi_PR_Err}), we
have\begin{equation}
1-2Q\left(\frac{\pi^{\frac{3}{2}}\left(\lambda_{0}\sqrt{\rho_{0}}+\lambda_{1}\sqrt{\rho_{1}}\right)}{\sqrt{2T_{0}}}\right)=\epsilon_{0}+\Delta\epsilon,\end{equation}
i.e., \begin{equation}
Q\left(\frac{\pi^{\frac{3}{2}}\left(\lambda_{0}\sqrt{\rho_{0}}+\lambda_{1}\sqrt{\rho_{1}}\right)}{\sqrt{2T_{0}}}\right)=\frac{1-\epsilon_{0}-\triangle\epsilon}{2}.\label{eq:Bi_PR_Err1}\end{equation}
 When $\epsilon_{0}\rightarrow0$ and $\Delta\epsilon\to0$, with
bivariate Taylor series expansion, we obtain
\begin{equation}
\frac{1}{2}-\frac{\pi\lambda_{0}\sqrt{\rho_{0}}}{2\sqrt{T_{0}}}-\frac{\pi\lambda_{1}\sqrt{\rho_{1}}}{2\sqrt{T_{0}}}=\frac{1-\epsilon_{0}-\triangle\epsilon}{2}.\label{eq:Bi_PR_Taylor}\end{equation}
If we choose $\lambda_{0}=\lambda_{0}^{\epsilon_{0}}$ as in
(\ref{eq:PRs_L0}), the maximum allowable value of $\lambda_{1}$
corresponding to a target outage probability increment
$\Delta\epsilon$ is given by
\begin{equation}
\lambda_{1}^{\Delta\epsilon}=\frac{1}{\pi}\left(\frac{T_{0}}{\rho_{1}}\right)^{\frac{1}{2}}\Delta\epsilon=\frac{1}{\pi}\left(\frac{\rho_{0}}{\rho_{1}}\cdot\frac{r_{0}^{-4}}{\beta_{0}}-\frac{\eta}{\rho_{1}}\right)^{\frac{1}{2}}\Delta\epsilon\label{eq:Bi_SR_Lambda},\end{equation}
and the transmission capacity of the PR network is given by
\begin{equation}\label{eq:Bi_PR_Capacity}
C_{0}^{\epsilon}=R_{0}\lambda_{0}^{\epsilon_0}\left(1-\epsilon_{0}-\Delta\epsilon\right).
\end{equation}
As shown in (\ref{eq:Bi_PR_Taylor}), when the SR network is
presented, the outage probability of the PR network can be
approximated by an affine function of $\lambda_0$ and $\lambda_1$
over asymptotically small $\epsilon_0$'s and $\Delta\epsilon_0$'s.

\subsection{Transmission Capacity of the SR Network}
Denote the outage probability of the SR network as $\epsilon_1$,
 the
outage probability of the SR network is given by
\begin{equation}
\mathcal{P}^{1}=\textrm{Prob}\left(\frac{\rho_{1}r_{1}^{-\alpha}}{\eta+{\displaystyle
\sum_{i\in\Pi_{0}}\rho_{0}|X_{i}|^{-\alpha}}+{\displaystyle
\sum_{j\in\Pi_{1}}\rho_{1}|Y_{j}|^{-\alpha}}}\leq\beta_{1}\right){\displaystyle
=\epsilon_{1}}.\label{eq:Bi_SR_err}\end{equation}
 Rewrite (\ref{eq:Bi_SR_err}) as \begin{equation}
\textrm{Prob}\left(Z\geq\rho_{1}\frac{r_{1}^{-\alpha}}{\beta_{1}}-\eta\right){\displaystyle
=\epsilon_{1}}.\end{equation}
 Define $T_{1}=\rho_{1}\frac{r_{1}^{-\alpha}}{\beta_{1}}-\eta$, and we
have \begin{equation} \textrm{Prob}\left(Z\geq
T_{1}\right){\displaystyle=\epsilon_{1}}.\end{equation}
 Similar to (\ref{eq:Bi_PR_Err1}), we obtain \begin{equation}
Q\left(\frac{\pi^{\frac{3}{2}}\left(\lambda_{0}\sqrt{\rho_{0}}+\lambda_{1}\sqrt{\rho_{1}}\right)}{\sqrt{2T_{1}}}\right)=\frac{1-\epsilon_{1}}{2}.\end{equation}
 When $\epsilon_{1}\to0$, with bivariate Taylor series expansion, we
have
\begin{equation}
\frac{1}{2}-\frac{\pi\lambda_{0}\sqrt{\rho_{0}}}{2\sqrt{T_{1}}}-\frac{\pi\lambda_{1}\sqrt{\rho_{1}}}{2\sqrt{T_{1}}}=\frac{1-\epsilon_{1}}{2}.\label{eq:Bi_SR_err1}\end{equation}
Therefore, the outage probability of the SR network is given by
\begin{equation}\label{eq:Bi_SR_E1}
\epsilon_{1}=\frac{\pi}{\sqrt{T_1}}\left(\lambda_{0}\sqrt{\rho_0}+\lambda_{1}\sqrt{\rho_1}\right),
\end{equation}
and the transmission capacity of the SR network is given by
\begin{equation}\label{eq:Bi_SR_Capacity}
C_{1}^{\epsilon}=R_{1}\lambda_{1}^{\epsilon}\left(1-\epsilon_{1}\right),
\end{equation}
where $R_{1}$ is the data rate adopted by successful SR links.

On the other hand, if we set the target outage probability of the PR
network to be $\epsilon_0+\Delta\epsilon$, and set the target outage
probability of the SR network to be $\epsilon_1$ simultaneously, we
could choose the value of $\lambda_{1}^{\epsilon}$ in
(\ref{eq:Bi_SR_Capacity}) as follows
\begin{equation}\label{eq:Bi_constr}
\lambda_{1}^{\epsilon}=\textrm{min}\left(\lambda_{1}^{\Delta\epsilon},\lambda_{1}^{\epsilon_1}\right),
\end{equation}
where $\lambda_{1}^{\epsilon_1}$ is given by (via
(\ref{eq:Bi_SR_E1}))
\begin{equation}\label{eq:Lambda1_E1}
\lambda_{1}^{\epsilon_{1}}=\frac{\epsilon_{1}}{\pi}\left(\frac{r_{1}^{-\alpha}}{\beta_{1}}-\frac{\eta}{\rho_1}\right)^{\frac{1}{2}}-\lambda_{0}^{\epsilon_{0}}\sqrt{\frac{\rho_0}{\rho_1}}.
\end{equation}

\subsection{Sum Transmission Capacity of the Overlaid Network}
When the SR network is present, based on the above analyses, the sum
transmission capacity of the overlaid networks is given by
\begin{eqnarray}
C_{s}^{\epsilon} &=& C_{0}^{\epsilon}+C_{1}^{\epsilon} \nonumber \\
&=&
R_{0}\lambda_{0}^{\epsilon_0}(1-\epsilon_{0}-\Delta\epsilon)+R_{1}\lambda_{1}^{\epsilon}(1-\epsilon_{1})\nonumber\\
&=&R_{0}\lambda_{0}^{\epsilon_0}\left(1-\frac{\pi}{\sqrt{T_0}}\left(\lambda_{0}^{\epsilon_0}\sqrt{\rho_0}+\lambda_{1}^{\epsilon}\sqrt{\rho_1}\right)\right)
\nonumber \\
&&{}+R_{1}\lambda_{1}^{\epsilon}\left(1-\frac{\pi}{\sqrt{T_1}}\left(\lambda_{0}^{\epsilon_0}\sqrt{\rho_0}+\lambda_{1}^{\epsilon}\sqrt{\rho_1}\right)\right)
\nonumber \\
&=&(R_{0}\lambda_{0}^{\epsilon_0}+R_{1}\lambda_{1}^{\epsilon})-\pi\left(\lambda_{0}^{\epsilon_0}\sqrt{\rho_0}+\lambda_{1}^{\epsilon}\sqrt{\rho_1}\right)\nonumber\\
&&{}\times\left(\frac{R_0}{\sqrt{T_0}}\lambda_{0}^{\epsilon_0}+\frac{R_1}{\sqrt{T_1}}\lambda_{1}^{\epsilon}\right).\label{eq:Capacity_Overlaid}
\end{eqnarray}
Compared to the single network case, the gain of the transmission
capacity (i.e., the overall spectrum efficiency) of the overlaid
networks over that of a single network is given by
\begin{equation} \label{eq:Kg}
K_g =
\frac{C_{s}^{\epsilon}}{C_{0}^{\epsilon_{0}}} \approx 1+\frac{C_{1}^{\epsilon}}{C_{0}^{\epsilon}}.
\end{equation}

\subsection{Tradeoff of the Transmission Capacities}
Here we consider two setups to study the tradeoff between the
transmission capacities of the PR network and the SR network. The
first setup is that we change the value of $\Delta\epsilon$ only,
and fix other parameters ($\rho_0$, $\rho_1$, $r_0$, $r_1$,
$\beta_0$, $\beta_1$, $\eta$, and $\epsilon_0$). The second setup is
that we change the value of $\rho_1$, and let other parameters
($\rho_0$, $r_0$, $r_1$, $\beta_0$, $\beta_1$, $\eta$, $\epsilon_0$,
and $\lambda_1$) be fixed.

Let us consider the first setup. When $\epsilon_0$ is fixed,
$\lambda_0$ is also fixed, see (\ref{eq:PRs_L0}). From
(\ref{eq:Bi_PR_Capacity}), we can see that $C_{0}^{\epsilon}$ is a
linear function of $\Delta\epsilon$. As such, when $\Delta\epsilon$
is increased, $C_{0}^{\epsilon}$ is reduced. Rewrite
(\ref{eq:Bi_SR_Capacity}) as
\begin{equation} \label{eq:C1_Tradeoff0}
C_{1}^{\epsilon}=\frac{R_1}{\pi}\sqrt{\frac{T_0}{\rho_1}}\Delta\epsilon\left(1-\sqrt{\frac{T_0}{T_1}}\epsilon_0-\sqrt{\frac{T_0}{T_1}}\Delta\epsilon\right).
\end{equation}
From (\ref{eq:C1_Tradeoff0}), we can easily verify that when
$\sqrt{T_1/T_0} > \epsilon_0$, $C_{1}^{\epsilon}$ is a convex
function of $\Delta\epsilon$, and when $\Delta\epsilon <
\frac{1}{2}(\sqrt{T_1/T_0}-\epsilon_0)$, $C_{1}^{\epsilon}$
increases monotonically over $\Delta\epsilon$.

Now, we consider the second setup. Rewrite (\ref{eq:Bi_PR_Capacity})
and (\ref{eq:Bi_SR_Capacity}) as follows,
\begin{equation}\label{eq:traeoff_C0}
C_{0}^{\epsilon}=R_{0}\lambda_{0}^{\epsilon_0}(1-\epsilon_{0}-\frac{\pi}{\sqrt{T_0}}\lambda_{1}^{\epsilon}\sqrt{\rho_1})
\end{equation}
and
\begin{equation}\label{eq:tradeoff_C1}
C_{1}^{\epsilon}=R_{1}\lambda_{1}^{\epsilon}\left(1-\frac{\pi\lambda_{0}^{\epsilon_0}}{\sqrt{\frac{\rho_1}{\rho_0}\frac{r_{1}^{-\alpha}}{\beta_1}-\frac{\eta}{\rho_0}}}+\frac{\pi\lambda_{1}^{\epsilon}}{\sqrt{\frac{r_{1}^{-\alpha}}{\beta_1}-\frac{\eta}{\rho_1}}}\right).
\end{equation}
We can easily show that when $\rho_1$ increases, $C_{0}^{\epsilon}$
decreases and $C_{1}^{\epsilon}$ increases.

\section{Numerical Results and Interpretations}
In this section, we present some numerical results based on our
previous analyses and give some interpretations. We set the values
of the network parameters as in Table~\ref{Tb:Pars} unless otherwise
specified.
\begin{table}
  \centering
  \caption{Network Parameters.}\label{Tb:Pars}
  \begin{tabular}{|c|c|c|}
    \hline
    Symbol & Description & Value \\
    \hline
    $\rho_0$ & Transmission power of PR TXs & 20 \textrm{W}\\
    \hline
    $\rho_1$ & Transmission power of SR TXs & 0.1 \textrm{W}\\
    \hline
    $r_0$ & Transmission range of PR TXs & 20 \textrm{m}\\
    \hline
    $r_1$ & Transmission range of SR TXs & 5 \textrm{m}\\
    \hline
    $\eta$ & Average power of ambient noise & $10^{-6}$ \textrm{W} \\
    \hline
    $\beta_0$ & Target SINR for PR network & 10 \textrm{dB} \\
    \hline
    $\beta_1$ & Target SINR for SR network & 10 \textrm{dB} \\
    \hline
  \end{tabular}
\end{table}

\subsection{Single Network Case}

\begin{figure}[htb]
\centerline{\scalebox{0.40}{
\input{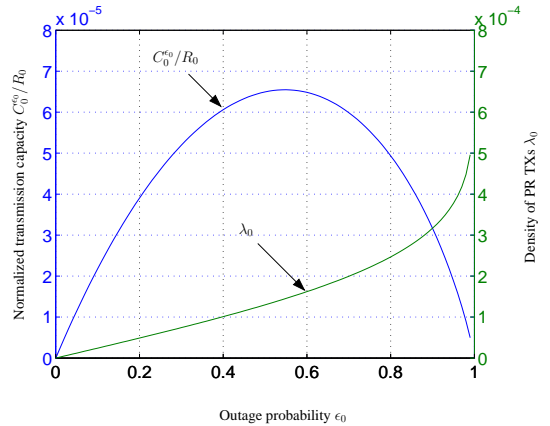}
}} \caption{Normalized transmission capacity/density of PR TXs vs.
outage probability for the PR network when the SR network is
absent.} \label{Fig:Tc_L0_PR0}
\end{figure}

In Fig.~\ref{Fig:Tc_L0_PR0}, we show the normalized transmission
capacity $C_{0}^{\epsilon_0}/R_0$ as a function of the outage
probability $\epsilon_{0}$, as well as the density of PR TXs
$\lambda_0$ vs. the outage probability $\epsilon_{0}$. Note that
these are exact results (not asymptotic ones) by using
(\ref{eq:Transmission_capacity}) and (\ref{eq:Qfunc_PR}). We could
see from this figure that when $\epsilon_0$ is about 0.55,
$C_{0}^{\epsilon_0}$ is maximized, and when $\epsilon_0<0.4$,
$\lambda_0$ is nearly a linear function of $\epsilon_0$, which
verifies the asymptotic result in (\ref{eq:PRs_L0}).

\begin{figure}[htb]
\centerline{ \scalebox{0.45}{
\input{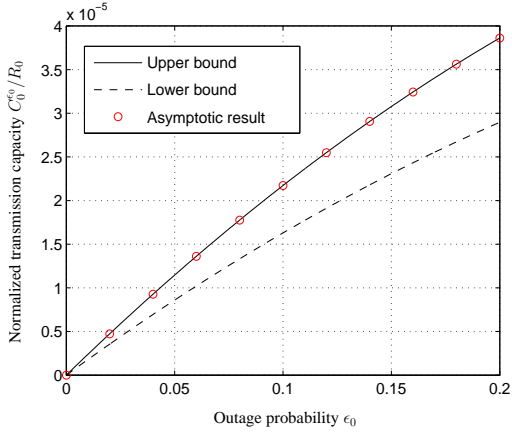}
} } \caption{Normalized transmission capacity vs. outage probability
for the PR network when SR network is absent.} \label{Fig:TC_PR0}
\end{figure}

In Fig.~\ref{Fig:TC_PR0}, we show the normalized asymptotic
transmission capacity $C_{0}^{\epsilon_0}/R_{0}$ as a function of
the outage probability $\epsilon_0$, and the upper and lower bounds
of the transmission capacity based on the results derived
in~\cite{Weber:Capacity}, which verifies the tightness of the upper
bound.

\subsection{Overlaid Network Case}
\begin{figure}[htb]
\centerline{\scalebox{0.45}{
\input{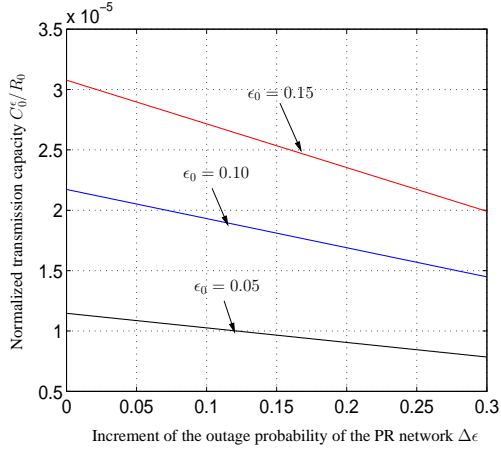}
}} \caption{Normalized transmission capacity of the PR network vs.
increment of the outage probability of the PR network.}
\label{fig:Outage_PR}
\end{figure}

The normalized transmission capacity of the PR network
$C_{0}^\epsilon/R_{0}$ vs. the increment of the outage probability
$\Delta\epsilon$ of the PR network is shown in
Fig.~\ref{fig:Outage_PR}. As expected, $C_{0}^\epsilon/R_{0}$ is
inversely proportional to $\Delta\epsilon$. On the other hand, since
$C_{0}^{\epsilon}$ is a convex function of $\epsilon_0$; and when
$\epsilon_0<\frac{1-\Delta\epsilon}{2}$, $C_{0}^{\epsilon}$
increases over $\epsilon_{0}$ monotonically for a fixed
$\Delta\epsilon$.

In Fig.~4, we show the normalized transmission capacity of the SR
network $C_{1}^{\epsilon}/R_1$ as a function of $\Delta\epsilon$,
see (\ref{eq:Bi_SR_Capacity}). As shown in the figure, we see that
$C_{1}^{\epsilon}$ increases monotonically over $\Delta\epsilon$,
since the larger $\Delta\epsilon$ is, the larger the values of
$\lambda_{1}^{\epsilon}$ and $\epsilon_1$ are, but the effect of
$\lambda_{1}^{\epsilon}$ on $C_{1}^{\epsilon}$ is dominant when
$\epsilon_{1}$ is small.
\begin{figure}[htb]
\centerline{\scalebox{0.42}{
\input{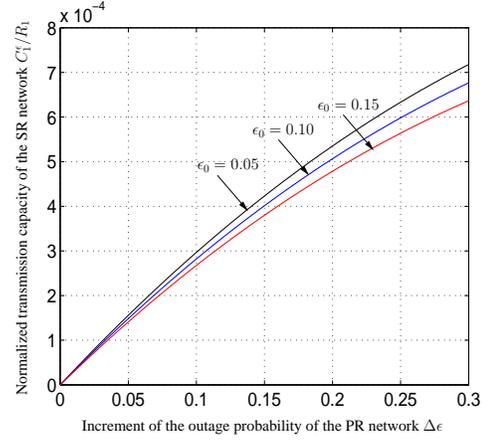}
}} \caption{Normalized transmission capacity of the SR network vs.
increment of the outage probability of the PR network.}
\label{fig:Outage_SR0}
\end{figure}

\begin{figure}[htb]
\centerline{\scalebox{0.45}{
\input{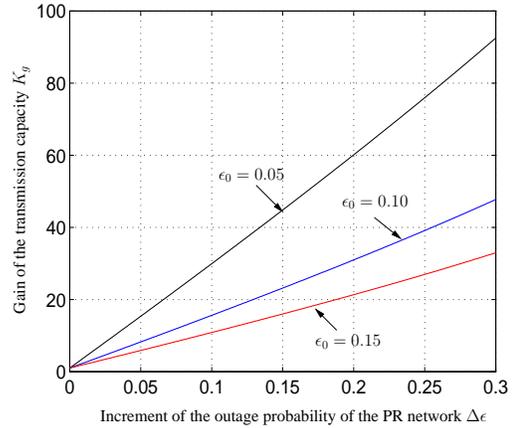}
}} \caption{Gain of the transmission capacity of the overlaid
network over that of the PR network.} \label{fig:Tc_gain}
\end{figure}

Assuming that $R_{0}=R_{1}$, the capacity gain $K_g$ of the overlaid
networks (i.e., the sum transmission capacity) over that of a single
network is shown in Fig.~\ref{fig:Tc_gain}, see (\ref{eq:Kg}). We
see that $K_g$ increases over $\Delta\epsilon$ since the extra
capacity contribution from the secondary network increases over
$\Delta\epsilon$.

\begin{figure}[htb]
\centerline{\scalebox{0.46}{
\input{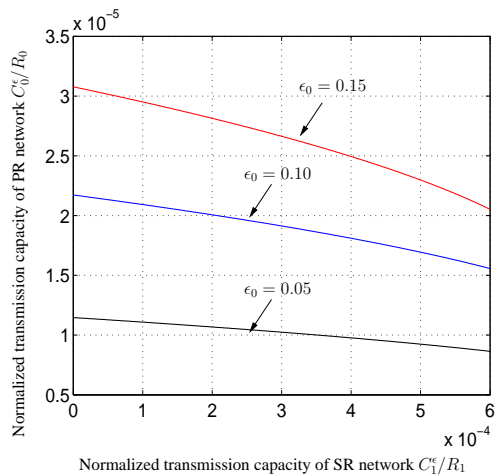}
}}\caption{Tradeoff of the transmission capacities of the PR and the
SR networks when the value of $\Delta\epsilon$ is changed.}
\label{fig:C0_vs_C1}
\end{figure}

In Fig.~\ref{fig:C0_vs_C1}, we show the tradeoff between the
normalized transmission capacity of the PR network
$C_{0}^{\epsilon}/R_0$ and that of the SR network
$C_{1}^{\epsilon}/R_1$ when $\Delta\epsilon$ changes as an
intermediate variable. We see that $C_{0}^{\epsilon}$ decreases over
$C_{1}^{\epsilon}$, which verifies the result in Section IV.

\section{Conclusions}
In this paper, we extended the concept of transmission capacity
defined for the single network case to overlaid network case. By
considering the mutual interference effect across two overlaid
networks, i.e., the PR network vs. the SR network, we derived the
transmission capacities for these two networks and studied their
tradeoffs. Different from the previous approach for the single
network case, we resorted to obtain the asymptotic solutions for
these capacities. The results showed that by letting a SR network
coexist with a legacy PR network, the spectrum efficiency per unit
area could be increased significantly. Although we focused on a
simple path-loss channel model with single-hop transmissions, the
results are meaningful and motivating us to study more complex cases
in the future work.

\end{document}